ASTRONOMY

# Stellar feedback and triggered star formation in the prototypical bubble RCW 120

**Matteo Luisi[1,2]\*, Loren D. Anderson[1,2,3], Nicola Schneider[4], Robert Simon[4], Slawa Kabanovic[4], Rolf Güsten[5], Annie Zavagno[6], Patrick S. Broos[7], Christof Buchbender[4], Cristian Guevara[4], Karl Jacobs[4], Matthias Justen[4], Bernd Klein[5], Dylan Linville[1,2], Markus Röllig[4], Delphine Russeil[6], Jürgen Stutzki[4], Maitraiyee Tiwari[5,8], Leisa K. Townsley[7], Alexander G. G. M. Tielens[8,9]**

Radiative and mechanical feedback of massive stars regulates star formation and galaxy evolution. Positive feedback triggers the creation of new stars by collecting dense shells of gas, while negative feedback disrupts star formation by shredding molecular clouds. Although key to understanding star formation, their relative importance is unknown. Here, we report velocity-resolved observations from the SOFIA (Stratospheric Observatory for Infrared Astronomy) legacy program FEEDBACK of the massive star-forming region RCW 120 in the [CII] 1.9-THz fine-structure line, revealing a gas shell expanding at 15 km/s. Complementary APEX (Atacama Pathfinder Experiment) CO J = 3-2 345-GHz observations exhibit a ring structure of molecular gas, fragmented into clumps that are actively forming stars. Our observations demonstrate that triggered star formation can occur on much shorter time scales than hitherto thought (<0.15 million years), suggesting that positive feedback operates on short time periods.

## INTRODUCTION

Radiative and mechanical feedback by massive stars on their nascent clouds is thought to play a key role in limiting the star formation efficiency of molecular clouds ("negative feedback"), and this controls the evolution of galaxies (*1*–*3*). Photo-ionization by ultraviolet photons will set up evaporative ionized gas flows, removing material from these clouds (*4*). Massive stars also have powerful winds that efficiently remove material from molecular clouds (*5*–*8*). Both processes often create a bubble geometry, with the formation of a compressed shell in which new stars form ("positive feedback") when the gas becomes gravitationally unstable (*9*, *10*). While observations reveal that bubbles are common in regions of massive star formation (*11*, *12*), the relative importance of the effects of positive and negative feedback from high-mass stars on future star formation is unknown.

RCW 120 is a nearby (~1.7 kpc) HII region with a physical diameter of ~4.5 pc (*13*). The region is ionized by a single O8V star, CD −38°11636 (LSS 3959) (*14*), and is known for its near-perfect circular symmetry at mid-infrared (IR) wavelengths (*15*, *16*). The expansion of the HII region has presumably led to the formation of massive clumps along the boundary that are the sites of recent and ongoing star formation (*17*–*20*). RCW 120 is considered to be the prototypical triggered star formation source given the simple region morphology and the abundance of ongoing star formation around its edges. Because of its vicinity and its importance in the context of triggered star formation, RCW 120 has been studied extensively through observations (*15*, *16*, *18*, *19*, *21*–*24*) and simulations (*25*–*29*).

## RESULTS AND DISCUSSION

### Observations of RCW 120

We have used the upGREAT (German Receiver for Astronomy at Terahertz) instrument on the Stratospheric Observatory for Infrared Astronomy (SOFIA) to survey RCW 120 in the [CII] 1.9-THz fine-structure line of ionized carbon. These observations are part of the SOFIA FEEDBACK legacy program dedicated to study the interaction of massive stars with their environment (*30*). We have also used the Large APEX sub-Millimetre Array (LAsMA) instrument on the Atacama Pathfinder Experiment (APEX) to observe the J = 3-2 isotopologues of CO (see Materials and Methods).

The strongest [CII] emission toward RCW 120 is found in a limb-brightened ring surrounding the HII region, with lower integrated intensities toward the HII region interior (see Fig. 1). The [CII] emission is morphologically similar to the 8-μm emission, which is a reliable tracer of the photodissociation region (PDR), the interface between the fully ionized HII region, and the surrounding neutral and molecular gas. The $^{12}CO(3-2)$ and $^{13}CO(3-2)$ emission around RCW 120 is found in denser clumps and filaments compared to that of [CII]. Although we have not observed the northwestern quadrant of the region in [CII], our CO measurements indicate that the ring is open toward the north and shows evidence for punctures in several locations.

Kinematic analysis of the [CII] data reveals a blue-shifted shell, which is visible as a curved structure in position-velocity (p-v) diagrams (see Fig. 2). The morphology of the emission is roughly isotropic, consistent with that of a half-shell expanding toward us at ~15 km/s. Expansion signatures are absent in $^{12}CO(3-2)$ and $^{13}CO(3-2)$ emission. This implies that the $C^+/C/CO$ transition lies in the ambient molecular cloud beyond the shell. We also detect faint, but widespread, [CII] emission from the bubble interior at the systemic velocity of the large-scale cloud [−7.5 km/s, as determined from the APEX CO(3-2) data and Mopra CO(1-0) data (*18*)] and also, at some positions, at positive (red-shifted) velocities. The red-shifted emission indicates expansion with a velocity of ~10 km/s away from us with respect to the systemic velocity of the cloud (see the Supplementary Materials). While this signature is blended with that of the systemic [CII] emission in the p-v diagrams, it is seen as a separate component in a number of individual spectra toward the region. The red-shifted shell is less massive and more clumpy than the blue-shifted shell and may reflect localized punctures through the molecular gas found at the backside of RCW 120.

[1]Department of Physics and Astronomy, West Virginia University, Morgantown, WV 26506, USA. [2]Center for Gravitational Waves and Cosmology, West Virginia University, Chestnut Ridge Research Building, Morgantown, WV 26505, USA. [3]Adjunct Astronomer at the Green Bank Observatory, P.O. Box 2, Green Bank, WV 24944, USA. [4]I. Physik. Institut, University of Cologne, Zülpicher Str. 77, 50937 Cologne, Germany. [5]Max-Planck Institut für Radioastronomie, Auf dem Hügel 69, 53121 Bonn, Germany. [6]Aix Marseille Université, CNRS, CNES, LAM, Marseille, France. [7]Department of Astronomy and Astrophysics, 525 Davey Laboratory, Pennsylvania State University, University Park, PA 16802, USA. [8]Department of Astronomy, University of Maryland, College Park, MD 20742, USA. [9]Leiden Observatory, Leiden University, PO Box 9513, 2300 RA Leiden, Netherlands.
\*Corresponding author. Email: maluisi@mix.wvu.edu

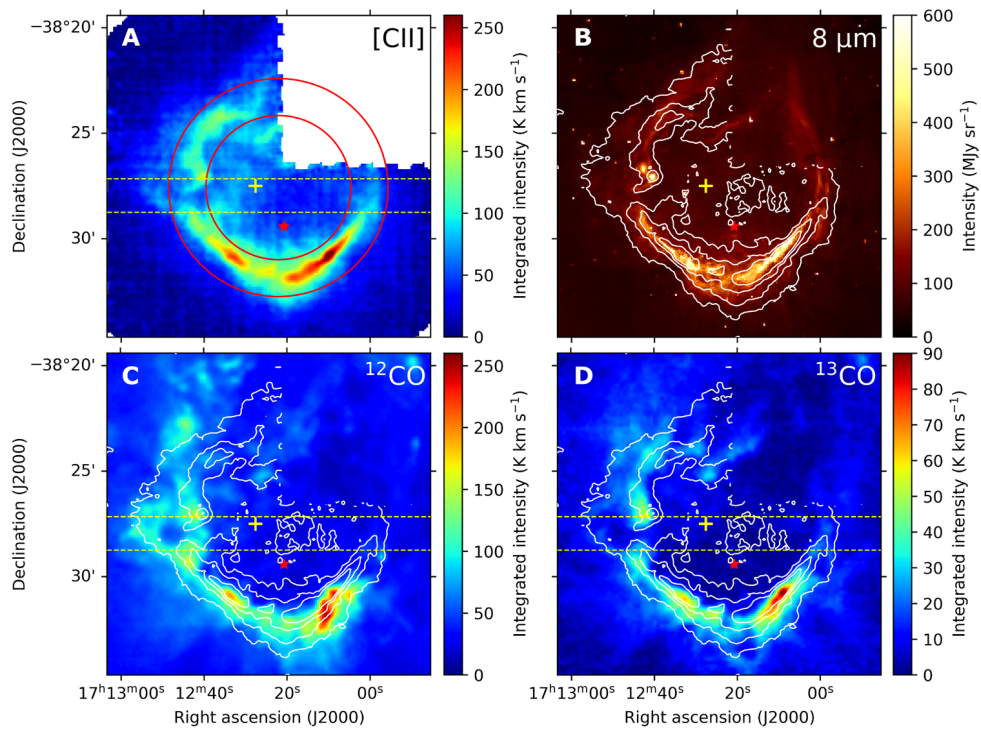

**Fig. 1. Morphology of RCW 120 in different tracers.** (**A**) SOFIA [CII] integrated intensity, scaled from 0 to 260 K km/s. The red circles indicate the approximate inner and outer PDR boundaries defined from Spitzer GLIMPSE 8-μm emission (*55*), and the red star shows the location of the ionizing source, CD −38°11636. The yellow "+" indicates "Position 1" (see the Supplementary Materials). (**B**) Spitzer GLIMPSE 8-μm emission. The contours are of [CII] integrated intensity, scaled from 40 to 160 K km/s in 40 K km/s increments. (**C** and **D**) APEX $^{12}$CO(3-2) and $^{13}$CO(3-2) integrated intensity, scaled from 0 to 260 K and 0 to 90 K km/s, respectively. The contours are the same as in (B). The areas enclosed by the dashed yellow lines in (A), (C), and (D) were used to extract the position-velocity diagrams shown in Fig. 2.

Given the expansion velocity of the [CII] shell, we estimate the dynamical age of the region to be ~0.15 million years (Ma), lower than previous estimates (*15, 20*), but roughly consistent with radiation-hydrodynamics simulations (*31*). The morphology of RCW 120 suggests that, with the exception of the background cloud at the systemic velocity, the entire shell is expanding at ~15 km/s.

### Diffuse x-ray emission

The Chandra x-ray observatory has measured diffuse x-ray emission toward RCW 120, revealing the existence of hot ($4 \times 10^6$ K) plasma created by the stellar wind from CD −38°11636 (*32*); see Fig. 3. Toward the dense southern shell, the hot plasma is confined to the interior of the bubble, but it has ruptured the shell toward the east and the north. [SII] observations show that the $10^4$ K HII region gas is primarily found toward the edges of the region (*33*). We find a filling factor of ~0.2 for the HII region gas (see the Supplementary Materials), suggesting that the remaining ~80% of the bubble is filled with the hot x-ray plasma.

### Stellar wind feedback and energetics

The picture that emerges from the wealth of our data reveals that the ~2500 km/s (*34*) wind from CD −38°11636 is shocked, creating a hot plasma that radiates in x-rays (Figs. 3 and 4). The overpressurized hot gas drives a strong shock wave, sweeping up the surrounding medium into a dense shell (*6*). This shell is ionized on the inside by extreme ultraviolet ($E > 13.6$ eV) photons, radiating in Hα and [SII] (*15, 33*). The remainder of the shell consists of neutral gas in a PDR created by impinging far ultraviolet ($6 < E < 13.6$ eV) photons (*35*). The PDR is traced by the 8-μm emission of polycyclic aromatic hydrocarbon molecules and by its dominant cooling line, the [CII] 1.9 THz fine-structure line (*36*); see Fig. 1.

The total kinetic energy of the expanding shell is $1 \times 10^{47}$ to $13 \times 10^{47}$ erg, which is a considerable fraction of the mechanical luminosity of the stellar wind over the age of the bubble ($15 \times 10^{47}$ erg). By comparison, the thermal energies of the ionized gas and the hot x-ray plasma are $5 \times 10^{46}$ and $17 \times 10^{46}$ erg, respectively (see Table 1). The ratio of the kinetic energy of the swept up shell to the thermal energy of the hot plasma is possibly much larger (up to a factor of ~8) than predicted for adiabatic expansion of stellar wind bubbles of 1.2 (*6*). The thermal energy may be reduced because of the breach in the northeast, where hot plasma appears to be leaking out of the HII region. Hot gas is also flowing out through the northern opening of the shell (Fig. 1), which may have further reduced the thermal energy of the plasma. The x-ray emission from the plasma in the north is likely obscured by a higher column density of neutral foreground gas. Furthermore, this ratio does not take into account electron conduction of the hot gas into the shell balanced by mass evaporation of shell gas into the hot gas and rapid radiative cooling in the mixing layer. This mass loading dissipates additional thermal energy, while simultaneously decreasing the pressure in the bubble. While the leaking hot gas may have stopped to inject energy into the expanding shell, the breaches in the northeast and north may have occurred recently enough that this would have only a small effect on the kinetic energy of the shell. We find a thermal pressure of $3 \times 10^4$ K cm$^{-3}$ in the surrounding molecular cloud in which the shell is expanding, and a ram pressure from the stellar wind of $8 \times 10^5$ K cm$^{-3}$, which is similar to the measured thermal pressure of the hot x-ray plasma of $7 \times 10^5$ K cm$^{-3}$.

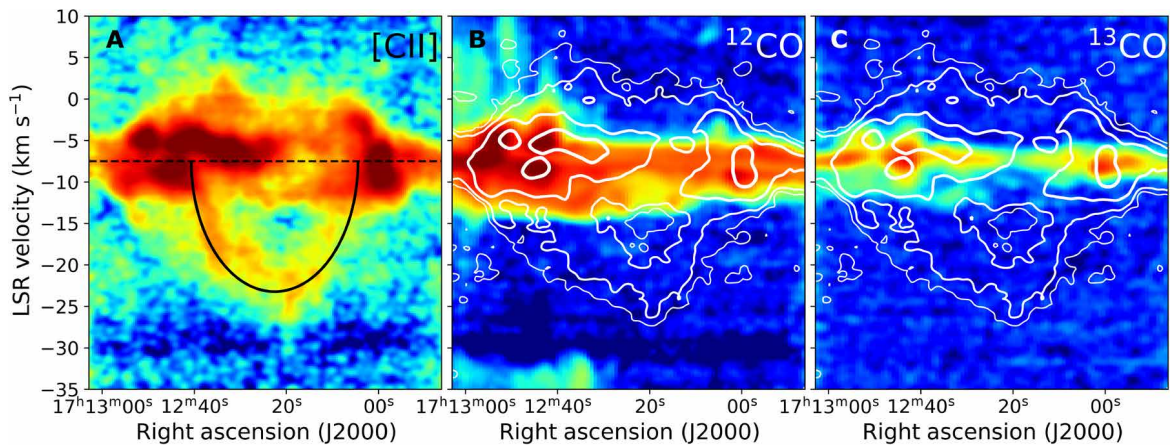

**Fig. 2. Position-velocity (p-v) diagrams.** (**A**) p-v diagram of the [CII] emission extracted from the area within the dashed yellow lines shown in Fig. 1A. The data were smoothed to a spatial resolution of 20″ and a velocity resolution of 0.8 km/s. The color scale is logarithmic and was chosen to highlight the blue-shifted, curved [CII] emission. The dashed black line shows the systemic [CII] velocity of the region and the black semi-ellipse shows our best-fit solution for the curve. The morphology of the blue-shifted emission reveals the presence of a [CII] shell expanding at ~15 km/s toward us. The red-shifted shell is not readily apparent in the p-v diagram as it is fainter and blended in with the bulk emission. (**B**) Same, for $^{12}$CO(3-2) emission. The white contours are of the [CII] emission shown in (A). (**C**) Same, for $^{13}$CO(3-2) emission. Expansion signatures are not seen in $^{12}$CO and $^{13}$CO.

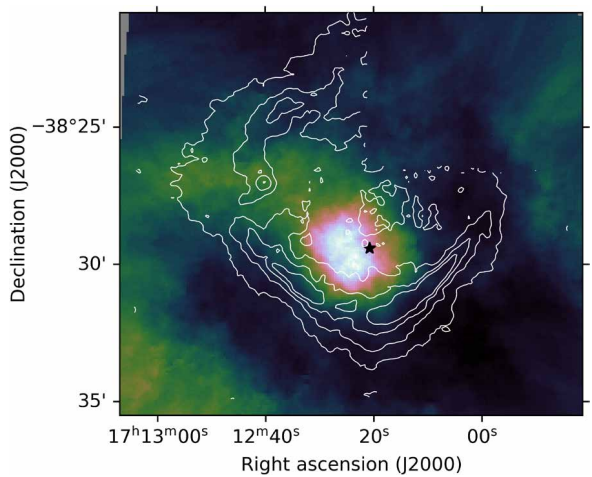

**Fig. 3. Chandra diffuse x-ray emission toward RCW 120.** The color scale ranges from 0 to $1.1 \times 10^{-9}$ photons cm$^{-2}$ s$^{-1}$ arc sec$^{-2}$. The contours are of integrated [CII] intensity, scaled from 40 to 160 K km/s in 40 K km/s increments, and the location of the ionizing source is marked by the black star. There is strong diffuse x-ray emission within the bubble, which is breaching the PDR toward the northeast.

In a system in which the ionizing star is moving supersonically with respect to the ambient cloud, a bow shock forms ahead of the star. The open shell toward the north of RCW 120 and the offset of the star from the center of the bubble are reminiscent of bow shock models for ultracompact HII regions such as G29.96-0.02 (*27*, *37–40*). In these models, a "stationary" shell structure (in the rest frame of the star) develops rapidly at the standoff distance of the bow shock that channels the flow of the material around it. The hot plasma, confined by the shell, is forced to flow downstream, where it quickly punches through the shell. A detailed model developed specifically for RCW 120 (*27*) explains the observed morphology well for a <~4 km/s motion of the star in the plane of the sky, consistent with its estimated proper motion of 2 to 4 km/s (*22*).

However, the mass flow around the shell in this model is at variance with the velocity field revealed by our observations. These models suggest that convection of energy and mass across the contact discontinuity between the hot plasma and the photo-ionized inner shell gas leads to rapid cooling and, as the thermal energy added by the wind is quickly lost, the expansion stalls. At that point, the main flow of the neutral gas is along the shell walls. Given the observed rapid expansion, we conclude that these models overestimate the energy and mass transfer across this boundary possibly because energy conduction by electrons is greatly limited by the magnetic field (*41*). As argued below, however, substantial magnetic fields are unlikely in RCW 120, and this effect possibly does not play a role in this source. In addition, the structure of the boundary in the model is determined by numerical diffusion, rather than by physical processes such as thermal conduction (*27*). We also see little evidence for Kelvin-Helmholtz instabilities that could mix cold shell gas into the hot plasma. The 24-μm dust arc observed at a distance of 0.25 pc from the star indicates a smooth flow of the ionized gas from the (southern) ionization front into the bubble and then on toward the north (*27*, *42*).

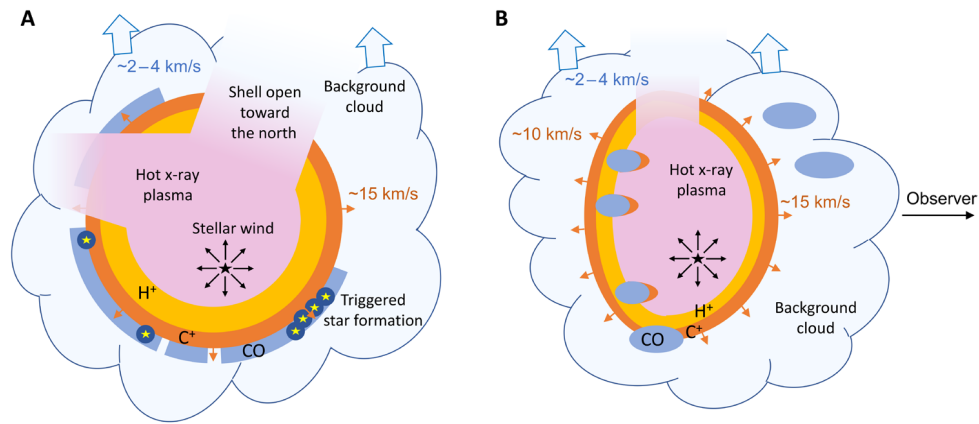

**Fig. 4. Sketch of the structure of RCW 120.** (**A**) Face-on view. The stellar wind from CD −38°11636 shocks the surrounding gas and creates a hot x-ray–emitting plasma. The energy injected by the stellar wind drives the rapid (~15 km/s) expansion of the [CII] shell. This expansion compresses the surrounding molecular gas and is responsible for the triggered star formation observed near the region edges. The shell is punctured toward the east and open toward the north, where the hot plasma is leaking into the surrounding medium. The surrounding large-scale molecular background cloud moves with a velocity of 2 to 4 km/s to the north with respect to the ionizing star (*22*, *27*). (**B**) Side view. The expansion of the shell toward its rear side is partially prevented by higher-pressure molecular gas structures. The systemic [CII] emission at a velocity of −7.5 km/s originates from these regions. The back side expands freely at several locations, giving rise to the red-shifted [CII] shell. Toward the front, the shell is expanding more homogeneously, suggesting that the foreground molecular gas is not strongly blocking the shell.

In addition, the morphology of the IR and x-ray observations reveal the presence of a puncture of the shell on the east side (Fig. 3). The surrounding cloud toward the rear side of the region has partially stopped the expansion of the shell in that direction. However, the [CII] velocity data suggest that the shell has also been punctured toward the rear side. This has allowed some of the x-ray emitting gas to flow out into the surrounding filamentary structure of the background cloud. The fact that these punctures predominantly occur toward the shell in the rear but not the front may be connected to instabilities caused by the interaction of the hot plasma within the bubble with the shell. There is no evidence that either the leakage or the surrounding cloud have stopped the expansion of the shell in any other direction. While traditional uniform background density models provide no explanation for such punctures, smoothed particle hydrodynamics models using an initial fractal density structure are able to reproduce perforations in the shell (*26*). This model, however, assumed that preexisting clumps are compressed by ionization erosion, for which we do not have evidence here. Further numerical simulations (see the Supplementary Materials) may help in determining the origin of these effects.

**Triggered star formation**
Morphologically, the star-forming CO clumps visible in Fig. 1 are part of the shell and are formed by compression in situ as preexisting clumps would be quickly left behind by the expanding shell inside the ionized gas/plasma bubble. Given the absence of pillars and globules in RCW 120, which are typically found in simulations with high fractal dimensions (*26*) and a high degree of turbulence (*43*), we believe that the initial molecular cloud was fairly homogeneous, making a substantial number of preexisting clumps unlikely. The densest clumps in the shell have already formed stars (see the Supplementary Materials). As a corollary, star formation must have proceeded very rapidly after shell formation. The shell expands supersonically into the background cloud, shocking the swept up gas at 15 km/s. The shocked gas will cool rapidly and that greatly increases the density. For a thin shell ($N_H < ~5 \times 10^{21}$ cm$^{-2}$), the [CII] fine-structure line will limit the cooling to about $T > ~100$ K in the PDR side facing the star, corresponding to a sound speed of $>~1$ km/s in the PDR. This implies a compression by a factor ($v_s/c_{PDR}$)$^2$ of $<~200$ (where $v_s$ is the speed of the shell and $c_{PDR}$ is the sound speed in the PDR). For larger column densities, CO can form in the PDR, cooling the gas down to ~20 to 30 K. At 30 K, the sound speed is ~0.5 km/s, and the resulting compression is then a factor of ~1000. With a preshock density of ~$10^3$ cm$^{-3}$ (see the Supplementary Materials), the clump density is estimated to be ~$10^6$ cm$^{-3}$. The fastest collapse time scale is the free-fall time scale, $t_{ff} = 0.06(10^6 \text{cm}^{-3}/n)^{1/2}$ Ma. Assuming a lower preshock density of $10^2$ cm$^{-3}$, the clump density would be $10^5$ cm$^{-3}$, corresponding to a free-fall time of 0.2 Ma. With an estimated lifetime of ~0.15 Ma for the bubble, the shell itself has not had the time to collapse, but the densest CO clumps would have had enough time to form stars after being compressed by the expanding shell. Several of the densest clumps have already formed stars, particularly in the more highly compressed shell toward the south of the region.

The absence of an expansion signature in $^{12}$CO(3-2) and $^{13}$CO(3-2) emission is at first glance unexpected as simulations qualitatively suggest that gas is "pushed away" during the expansion of the HII region (*26*, *43*). Although this may appear intuitive, models show that the difference between the initial velocity of the cloud and that of formed clumps after being compressed by an expanding shell is small (on the order of one to a few kilometers per second) (*44*). For the CO structures in the backside of the HII region, the expansion would naturally be much slower than that of the [CII] shell due to the large density variations between the clump and interclump gas, whereas any motion of the dense, star-forming clumps in the southern ring would be perpendicular to the line of sight and therefore not detected spectrally. These results are consistent with the observed $^{12}$CO(3-2) and $^{13}$CO(3-2) velocity structure and that of clumps formed by cooling and compression in the expanding shell.

In a future study, we intend to discuss the origin scenario of RCW 120, to investigate the hypothesis that the region was formed by the collision of two molecular clouds (*21*). In this scenario, the

Table 1. Properties of the different components of RCW 120

| Component | Mass ($M_\odot$) | Thermal energy ($10^{46}$ erg) | Kinetic energy ($10^{46}$ erg) | Luminosity ($L_\odot$) | Pressure ($10^6$ K cm$^{-3}$) |
|---|---|---|---|---|---|
| Expanding [CII] shell | 40–520* | 0.1–1.3 | 10–120 | - | 1–10 |
| Molecular gas | 2500 | 2.1 | - | - | 0.015 |
| Ionized gas | 26† | 5.1 | - | - | 8‡ |
| Stellar wind§ | - | - | 150 | 80‖ | 0.8¶ |
| Hot x-ray plasma | 0.05 | 17 | - | 0.10 | 0.7 |
| CD −38°11636 | - | - | - | $9.1 \times 10^4$ | - |
| [CII] | - | - | - | 690 | - |
| $^{12}$CO(3-2) | - | - | - | 2.6 | - |
| $^{13}$CO(3-2) | - | - | - | 0.58 | - |

*Total atomic mass. †From [SII] observations (33). ‡From (20). §Over the lifetime of the bubble. ‖Mechanical luminosity. ¶Dynamical pressure.

limb-brightened ring seen in the IR is interpreted as a cavity created by the cloud collision. Here, we emphasize that the CO data do not probe the expansion of the HII region but rather indicate the presence of two distinct clouds (see the Supplementary Materials). The observed morphology and velocity signature of the [CII] emission, however, reveal a relatively homogeneously expanding shell in RCW 120, making it unlikely that the shell structure itself was directly produced by the collision of two molecular clouds. This is very different from the study of the Infrared Dark Cloud G035.39-00.33, whose emission is interpreted in terms of colliding clouds (45). More work is required to fully understand the formation mechanism of RCW 120 in the context of the rapidly expanding shell seen in [CII].

The short lifetime of the bubble suggests that magnetic fields do not play a large role in its evolution. The high overpressure of the hot plasma and the ionized gas ensures that the global dynamics of the bubble are not affected by the magnetic field of the molecular cloud (31). If magnetic field strengths were substantial, the magnetic field would be frozen in the compressed shell and this would cushion the shock, limiting its compression efficiency (46). Models show that magnetic fields are, in general, preferentially oriented parallel to the expansion front (31, 47). In this case, fragmentation would not be prevented, and the fragmentation time scale would be comparable to the free-fall time. Magnetically supercritical cores would be able to form and ambipolar diffusion, which typically works on time scales several times the free-fall time scale (47, 48), may be minor. However, further gravitational collapse of fragments is likely to take longer due to the strong magnetic support. Further magnetohydrodynamic studies, including self-gravity, will be required to determine the influence of magnetic fields on the collapse of wind-driven shells.

We note that the Veil nebula in Orion—the shell driven by the stellar wind from $\theta^1$ Ori C—shows no evidence for triggered star formation (7, 8). This may reflect its limited column density and the absence of CO condensations (49). Alternatively, these differences may result from the presence of a substantial magnetic field in the Veil of $|B| > \sim 50$ μG (50).

Half of all Galactic massive star formation regions create IR bubbles (12). If rapid expansion is a general characteristic of star-forming regions, our results have major ramifications for the study of galaxy formation in the early universe and their subsequent evolution. Little is known about the formation mechanisms of stars in the early universe. As the environment then was metal-poor, the stars are expected to have been massive and therefore would have produced strong feedback effects. Our observations suggest that rapid triggered star formation around these massive stars may have played an important role, possibly contributing to the large star formation densities observed in high-redshift galaxies (51). A broader survey of massive star formation regions in [CII] emission, as is currently undertaken in the SOFIA FEEDBACK legacy program (30), is required to understand the role of massive stars in the collection, compression, fragmentation, and collapse of molecular clouds.

## MATERIALS AND METHODS

The [CII] line at 1.9 THz was observed with SOFIA on 10 June 2019 on a flight from Christchurch, New Zealand, using upGREAT (52). The map was split into four tiles (each covering an area of 7.26′ × 7.26′), of which three were observed. Each tile was mapped four times in total-power array-on-the-fly mode. The first two coverages were done once horizontally (along right ascension) and once vertically (along declination), with the array rotated by 19° on the sky. The second two coverages were shifted by 36″ to achieve the best possible coverage for the [OI] line (which was observed in parallel but is not discussed here). The final map is centered on Right Ascension (RA), Declination (Dec) [J2000] = $17^h12^m03.65^s$, −38°30′28.43″. A fast Fourier transform spectrometer with 4 GHz instantaneous bandwidth was used as the backend (53). The velocity resolution of the final data is 0.2 km/s. The intensity is given in units of main beam brightness ($T_{MB}$), assuming a forward efficiency, $\eta_f = 0.97$, and an average main beam efficiency of 0.65. The half-power beam width at 1.9 THz is 14.1″ and the final data have a pixel size of 5″. The root mean square (rms) main beam brightness is 1.8 K for 5″ pixels and 0.2 km/s velocity channels. The scientific objectives and technical details of the FEEDBACK program can be found at https://feedback.astro.umd.edu/ (30).

To improve the quality of the spectra, we identify systematic variations of the baseline originating in instabilities in the telescope system (namely from the backend, receiver, and telescope optics) and atmosphere over the course of the observations. During the calibration process, we produce an additional data product from the emission-free background source (OFF-source) measurements by subtracting subsequent OFF positions from each other. These "OFF-OFF" spectra are calibrated the same way as the "ON-OFF" spectra containing the astronomical information. The created set of spectra contains the dynamics of the telescope system and the atmosphere but no emission from the astronomical source. Subsequently, we identify features that account for most of the variance from the mean over the OFF-OFF spectra using principal component analysis (PCA). We do this individually for each of the 14 heterodyne pixels of the upGREAT low-frequency array (LFA) receiver and for each SOFIA flight where RCW 120 was observed. The PCA produces a set of components that describe the variance in the spectral structure over the course of the observations from which the baseline artifacts originate. We then reconstruct each ON-OFF spectrum containing information on the astronomical source in terms of these components via a linear combination. Here, a component is more strongly weighted the better the component describes the structure of the spectrum. Because the components are derived from a different set of observations than the ON-OFF spectra, only features characteristic to both sets of spectra will be included in this reconstruction. We lastly subtract each component from the ON-OFF spectrum. This process allows the removal of complex artificial features from a spectrum, which cannot be removed with a standard polynomial baseline fit.

RCW 120 was also mapped on 21 September 2019 in the $^{13}$CO(3-2) and $^{12}$CO(3-2) transitions at 330.6 and 345.8 GHz using the LAsMA array on the APEX telescope (54). LAsMA is a 7-pixel single polarization heterodyne array that allows simultaneous observations of the two isotopomers in the upper ($^{12}$CO) and lower ($^{13}$CO) sideband of the receiver, respectively. The array is arranged in a hexagonal configuration around a central pixel with a spacing of about two beam widths ($\theta_{mb}$ = 18.2″ at 345.8 GHz) between the pixels. The backends are advanced fast Fourier transform spectrometers (53) with a bandwidth of 2 × 4 GHz and a native spectral resolution of 61 kHz (0.053 km/s at 345 GHz). The mapping was done in total power on-the-fly mode using the same central position as the SOFIA [CII] map (RA, Dec [J2000] = $17^h12^m03.65^s$, $-38°30′28.43″$) and a clean reference position at RA, Dec [J2000] = $17^h10^m41^s$, $-37°44′04″$. The mapped region of 15′ × 15′ was split into four tiles. Each tile was scanned in both right ascension and declination with a spacing of 9.1″ (half the beam size) and oversampling to 6″ in scanning direction, resulting in a uniformly sampled map with high fidelity. The beam width at 345.8 GHz is 18.2″ and the final data cubes are gridded to a pixel size of 9.1″. The rms main beam temperature noise of the data is 0.5 K when resampled to 0.1 km/s in velocity.

## SUPPLEMENTARY MATERIALS

Supplementary material for this article is available at http://advances.sciencemag.org/cgi/content/full/7/15/eabe9511/DC1

**Acknowledgments:** This study was based on observations made with the NASA/DLR Stratospheric Observatory for Infrared Astronomy (SOFIA). SOFIA is jointly operated by the Universities Space Research Association Inc. (USRA), under NASA contract NAS2-97001, and the Deutsches SOFIA Institut (DSI), under DLR contract 50 OK 0901 to the University of Stuttgart. upGREAT is a development by the MPI für Radioastronomie and the KOSMA/Universität zu Köln, in cooperation with the DLR Institut für Optische Sensorsysteme. APEX is a collaboration between the Max-Planck-Institut für Radioastronomie, Onsala Space Observatory (OSO), and the European Southern Observatory (ESO). We thank M. Sánchez-Cruces for helpful review of the [SII] observations of RCW 120. **Funding:** We acknowledge funding by the Bundesministerium für Wirtschaft und Energie through the grant FEEDBACK (50OR1916). N.S. and R.S. acknowledge support from the ANR and DFG grant "GENESIS" (ANR-16-CE92-0035-01/DFG1591/2-1). The GREAT team members of the I. Physik. Institut, University of Cologne, and Max-Planck Institut für Radioastronomie, Bonn, acknowledge funding from the DFG, project number SFB 956, projects A4, D2, and D3. L.D.A. acknowledges support from NSF grant AST1516021. A.Z. thanks the support from the Institut Universitaire de France. Financial support for the SOFIA Legacy Program, FEEDBACK, at the University of Maryland was provided by NASA through award SOF070077 issued by USRA. **Author contributions:** A.G.G.M.T. and N.S. conceived the FEEDBACK legacy survey, which motivated this study. M.L. and A.G.G.M.T. wrote the manuscript with input from all other authors. M.L. analyzed the data and generated the figures under L.D.A.'s guidance. R.S., C.B., J.S., K.J., M.J., and B.K. generated and processed the SOFIA [CII] observations. R.G. generated and processed the APEX CO observations. S.K., N.S., A.G.G.M.T., L.D.A., R.S., M.R., C.G., D.L., M.T., A.Z., and D.R. assisted on the analysis and interpretation of the [CII] and CO data. L.K.T. and P.S.B. analyzed and interpreted the Chandra x-ray data. All authors discussed the results. **Competing interests:** The authors declare that they have no competing interests. **Data materials and availability:** The [CII] data used in this study are publicly available in the SOFIA Science Archive at IRSA (https://irsa.ipac.caltech.edu/). All data needed to evaluate the conclusions in the paper are present in the paper and/or the Supplementary Materials.

Submitted 24 September 2020
Accepted 19 February 2021
Published 9 April 2021
10.1126/sciadv.abe9511

Citation: M. Luisi, L. D. Anderson, N. Schneider, R. Simon, S. Kabanovic, R. Güsten, A. Zavagno, P. S. Broos, C. Buchbender, C. Guevara, K. Jacobs, M. Justen, B. Klein, D. Linville, M. Röllig, D. Russeil, J. Stutzki, M. Tiwari, L. K. Townsley, A. G. G. M. Tielens, Stellar feedback and triggered star formation in the prototypical bubble RCW 120. *Sci. Adv.* **7**, eabe9511 (2021).

Supplementary Materials for

**Stellar feedback and triggered star formation in the prototypical bubble RCW 120**

Matteo Luisi, Loren D. Anderson, Nicola Schneider, Robert Simon, Slawa Kabanovic, Rolf Güsten, Annie Zavagno, Patrick S. Broos, Christof Buchbender, Cristian Guevara, Karl Jacobs, Matthias Justen, Bernd Klein, Dylan Linville, Markus Röllig, Delphine Russeil, Jürgen Stutzki, Maitraiyee Tiwari, Leisa K. Townsley, Alexander G.G.M. Tielens

Correspondence to: maluisi@mix.wvu.edu

**This PDF file includes:**

Supplementary Text
Figs. S1 to S10
Caption for Movies S1 to S3

**Other Supplementary Materials for this manuscript include the following:**

Movies S1 to S3

**Supplementary Text**

Characteristics of RCW 120

The ionized gas of RCW 120 is surrounded by a dense limb-brightened ring of gas and dust. The dust has been observed by the *Herschel Space Observatory* at wavelengths of 70 μm to 500 μm *(17,19,56,57)*, by APEX-LABOCA at 870 μm *(16)*, and also by SEST-SIMBA at 1.2 mm *(15)*. Asymmetric column density profiles *(24)* and characteristic features in the column density probability density function of these far-IR observations *(25)* demonstrate the existence of a compressed neutral shell around RCW 120 caused by the expansion of the ionized gas into the surrounding molecular cloud. The molecular gas has been observed in CO(1-0), with no evidence for expansion of the molecular material *(18)*, but rather indicating the presence of two distinct clouds. The molecular clouds appear to be connected in velocity space toward the south of RCW 120, albeit several bubble radii away from the region, making it unlikely that this connection is due to the expansion of the ionized gas. It has been suggested that a collision between molecular clouds may have triggered the formation of the ionizing O star of RCW 120 *(21)*. The overall morphology of RCW 120 is very reminiscent of the interaction of a star with a stellar wind bubble moving with respect to the surrounding gas. Numerical simulations imply then that the ionizing star, CD -38°11636, has a proper motion at 2-4 km/s with respect to the background cloud, which may explain the offset of the star from the geometrical center of the bubble *(22,27)*. At an estimated temperature of 40 K (see below), the sound speed in the molecular cloud is 0.5 km/s. Thus, the star is estimated to move at a Mach number of 4 to 8 relative to the cloud while the shell expansion velocity of 15 km/s results in a shock with a Mach number of 30 relative to the unshocked molecular cloud material.

The dense ring surrounding RCW 120 hosts sites of ongoing star formation, including high-mass star formation *(58)*, which has often been explained as having been triggered by the expansion of the region, either by shell fragmentation *(24,25,59)* or by compressing pre-existing clumps *(26)*. Eight condensations in the field of RCW 120 were identified *(15)*, five of which are located in the dense ring. Numerous Class I and Class II young stellar objects (YSOs) are located in the ring (see Figure S1), including a massive Class 0 candidate coincident with the densest part of the ring *(16)*. Eleven of the YSOs are aligned parallel to the ionization front, equally spaced by ~0.1 pc, possibly as a result of Jeans gravitational instabilities.

Velocity structure of the gas

The [CII] velocity-resolved observations show the characteristic kinematics of a shell expanding toward us: at low velocities (~ -25 km/s) the emission is centralized, and as the velocity increases, a ring-shaped structure emerges and increases in radius until the systemic velocity of -7.5 km/s is reached. The systemic velocity was estimated by a Gaussian line fit to the APEX $^{13}$CO(3-2) data and can be seen in [CII] and CO(3-2) emission in Figures S2-S7 and Movies S1-S3. This estimate for the systemic velocity agrees with that found previously in CO(1-0) emission *(18)*. Red-shifted [CII] emission at ~2 km/s is found by examining spectra at several locations within the interior of the region, indicating expansion with a velocity of ~10 km/s away from us. The expanding shell is seen in [CII] spectra averaged over the region (Figure S8). Figure S8 also shows a typical [CII] spectrum toward the interior of the bubble to better identify the individual blue- and redshifted components that are partly diluted in the averaged spectrum. We do not detect a clear expansion signal in CO(3-2) emission (see Figures S3 and S4).

To find the expansion velocity of the shell, we perform an orthogonal distance regression on the intensity of the position-velocity (p-v) diagram (Figure 2), assuming that its shape can be expressed as a half-ellipse in p-v space. We leave the lengths of the semi-major and semi-minor axes, and the center Right Ascension of the half-ellipse as free parameters. To ensure convergence, however, we exclude emission that is clearly not associated with the shell, and fix the center velocity of the semi-ellipse at the systemic velocity of the region (-7.5 km/s). Our best-fit solution is consistent with an expansion velocity of 15.7 km/s.

We also estimate the expansion velocity by binning the data to 80" spatial resolution, removing a Gaussian fit of the [CII] emission at the systemic velocity from each spatial location, and finding the velocity in the residual spectrum that has the maximum intensity, after smoothing the Gaussian-subtracted spectra to 1 km/s (Figure S9). Overall, the blue-shifted velocities are similar to those of an isotropically expanding half-shell with a velocity of 14 km/s.

We assume an average expansion velocity of 15 km/s (the average of the two methods) for all further calculations. With a physical radius of the region of 2.25 pc (Figure 1), the dynamical age of the region is then ~0.15 Myr.

Physical conditions and energetics
In order to fully understand the processes involved in the expansion, we must quantify the masses and energetics associated with the expanding bubble. The [CII] line is affected by self-absorption toward the dense PDR in the ring. We therefore estimate the velocity-resolved optical depth from [$^{13}$CII] lines, which are shifted from that of [CII] by +11.2, -65.2, and +63.3 km/s to correct the intensity of the [CII] line and derive the column density *(60)*. The signal-to-noise ratio (SNR) of every single pixel is not large enough to disentangle the [$^{13}$CII] line from the noise floor. We therefore average over square-arcminute large regions to sufficiently increase the SNR. The averaged dense regions of the PDR are selected using a dendrogram-based technique *(61)*. We determine the local averaged column number density by scaling the [$^{13}$CII] line with the local carbon ratio $^{12}$C/$^{13}$C = 59 *(62)* and by assuming an excitation temperature of $T_{ex}$ = 100 K *(60)*. For locations less affected by self-absorption, we estimate the column density directly from the integrated [CII] line intensities.

We convert the [CII] column density to the hydrogen column density by applying the carbon-to-hydrogen ratio of $1.6 \times 10^{-4}$ *(63)*, assuming that all carbon in the shell is ionized. This calculation yields a total mass of 476 ± 20 M☉ for the annular PDR shown in Figure 1, where the uncertainty is determined by propagating the error of the distance to RCW 120 and the uncertainty in integrated [CII] intensity. We use an automated spectral decomposition pipeline based on Gausspy+ *(64)* in order to disentangle the blue- and red-shifted [CII] emission from that at the systemic velocity. The mass of the blue-shifted shell toward the interior of the region is 32 M☉, and the mass of the red-shifted emission is 9 M☉, summing up to ~40 M☉. This is the lower limit for the expanding shell mass, only determined from the emission clearly expanding along the line of sight. This is, however, unrealistic since at least a part of the ring emission, possibly even all emission, must also be associated with the expanding [CII] shell. We thus obtain an upper limit of 520 ± 20 M☉ for the shell mass, including the ring structure. With an average expansion speed of 15 km/s, the total kinetic energy of the shell is then $1-12 \times 10^{47}$ erg, and its thermal energy (assuming a temperature of 100 K) is 1-13×10$^{45}$ erg. Assuming an average density of 10$^4$-10$^5$ cm$^{-3}$, the thermal pressure of the shell is 1-10×10$^6$ K cm$^{-3}$.

The mass of the expanding shell is a significant fraction of the molecular gas mass of RCW 120. In order to derive the molecular gas mass *(65,66)*, we first estimate the excitation temperature, $T_{ex}$, from the $^{12}$CO(3-2) peak intensities ($T_{12CO}$):

$$T_{ex} = 16.6 \left[\ln\left(1 + \frac{16.6}{T_{12CO}+0.036}\right)\right]^{-1}.$$

We find an average $T_{ex}$=40 K toward the dense PDR, where $^{12}$CO(3-2) is optically thick. This value is similar to the derived dust temperature of RCW 120 *(19,20,67)*. The excitation temperature decreases to 15-20 K in the cool molecular gas outside of the PDR ring. Note that $T_{ex}$ is underestimated in regions of self-absorption. We then used the excitation temperature map to obtain a map of the $^{13}$CO optical depths, $\tau_{13CO}$, using

$$\tau_{13CO} = -\ln\left(1 - \frac{T_{13CO}}{15.8/(\exp(15.8/T_{ex})-1)-0.045}\right).$$

We find an average $^{13}$CO opacity of ~1 toward the PDR, and use this value for all further calculations. From the observed $^{13}$CO(3-2) integrated line intensities, $\int T_{13CO}\,dv$, the $^{13}$CO column density, $N_{13CO}$, is then

$$N_{13CO} = 5.29 \times 10^{12}(T_{ex} + 0.88)\exp\left(\frac{31.7}{T_{ex}}\right)\left(\frac{\tau_{13CO}}{1-\exp(-\tau_{13CO})}\right)\int T_{13CO}\,dv.$$

With a $^{12}$CO/$^{13}$CO abundance ratio of 59 *(62)*, and a $^{12}$CO/H$_2$ conversion factor of 8.5×10$^{-5}$ *(36)*, the total molecular gas mass in the ring is 2500 M⊙. Assuming a region radius of 2.25 pc, this mass corresponds to a mean molecular gas density of ~1100 cm$^{-3}$. Its thermal energy is of the order of 2.1×10$^{46}$ erg, and its thermal pressure is 1.5×10$^4$ K cm$^{-3}$.

For the ionized hydrogen gas, we adopt a mean density of $n_e$ = 22 cm$^{-3}$ *(33)*, resulting in an ionized gas mass of ~26 M⊙. From the Southern Galactic Plane Survey radio continuum data *(68)*, we find the emission measure (*EM*) averaged over all of RCW 120 to be ~1.2×10$^4$ pc cm$^{-6}$. The filling factor, *ff*, is then

$$ff = \left(\frac{EM}{n_e^2 l}\right)^{-1},$$

where the region diameter l=4.5 pc is the estimated path length. Using the above values, we find *ff* = 0.2. HII regions at the Galactocentric radius of RCW 120 have an average electron temperature of ~8000 K *(69)*, and therefore the thermal energy of the ionized gas of RCW 120 is 5×10$^{46}$ erg.

We derive the stellar wind mass loss rate, $\log(dM/dt) = -6.81$ M⊙/yr *(70)*, assuming standard luminosity, mass, and effective temperatures for an O8 star *(71)*, and a terminal-to-effective escape velocity ratio of 2.6. For O8 stars, stellar wind velocities of ~2500 km/s are typical *(34)*. The resulting mechanical energy injected by the stellar wind over the age of the region is then ~1.5×10$^{48}$ erg, corresponding to a mechanical luminosity of ~80 L⊙.

Using the *Chandra* X-ray data *(32)*, and a model of a lightly-absorbed soft plasma ($kT =$ ~0.35 $keV$), we calculate the thermal energy of the hot gas, $E_{th.}$. We find $E_{th} = 6.6 \times 10^{46}$ erg interior to the PDR, and $E_{th} = 12.5 \times 10^{46}$ erg toward the PDR breach in the north-east. The diffuse X-ray emission fills the center of the bubble and is brightest in the southeastern half of the bubble. We cannot exclude the possibility of hot gas in the northwestern half of the bubble, whose X-ray emission may be absorbed by a larger overlying column density. From the plasma model, the estimated density of the hot gas is ~0.05 cm$^{-3}$, and its pressure is ~$5 \times 10^5$ K cm$^{-3}$.

HII region and PDR conditions
We aim to investigate the physical conditions in the shell with a model of the HII region and a PDR model. There is some disagreement in the literature whether the [CII] emission toward HII regions traces primarily the neutral PDR, or if there is a significant distribution from the ionized plasma as well *(72-74)*. We use the spectral synthesis code Cloudy to estimate how much of the [CII] emission comes from the ionized phase, and how much comes from the PDR. The calculations were performed with version 17.02 of Cloudy *(75)*. We assume that the ionizing source is of type O8, with an effective temperature of 34,880 K and an ionizing photon (h$\nu$ > 13.6 eV) emission rate of $10^{48.44}$ s$^{-1}$ *(71)*. We further assume a radius of 2.25 pc for the region and a filling factor of 0.2 (see above). We find that the observed blue-shifted [CII] emission is well reproduced by a density profile of <~50 cm$^{-3}$ within the bubble, as suggested by the [SII] observations *(33)*, surrounded by a thin denser shell. The density and thickness of the shell cannot be well-constrained from our observational parameters, however, we find that only ~20% of the cumulative [CII] emission is due to the ionized HII region plasma. This supports our assumption that the red- and blue-shifted [CII] emission truly traces the expanding shell. Assuming typical HII region abundances, the electron temperature within the bubble is ~8000 K, and the average temperature of the shell is ~150 K.

We also analyze the physical properties of the molecular gas clumps found primarily toward the PDR to better understand their origin. In order to identify the clumps, we apply the clump finding algorithm GAUSSCLUMPS *(76)* to the $^{13}$CO(3-2) data cube. The algorithm decomposes the map into a number of clumps by assuming a Gaussian emission distribution in space and velocity for each clump *(77,78)*. The clump properties were then calculated according to the conversion formulas described above. We extracted ~100 clumps above the spatial and kinematic resolution limits with masses between 2 and 530 M$_\odot$, and radii up to 1.1 pc. Some of the most massive clumps correspond to the clumps detected in continuum *(17)*, although we do not expect a one-to-one correlation since we included kinematic information in our clump finding algorithm. The average clump density is typically a few $10^4$ cm$^{-3}$. Most of the clumps are pressure confined or transient which can be seen in Fig. S10, showing the virial parameter α, i.e. the ratio between virial mass and CO mass *(79)*. The virial mass is calculated by $M_{vir} = 378R(\Delta v)^2$, where $R$ is the clump radius and $\Delta v$ is the CO line width *(80)*. However, there are a number of clumps, mostly located in the PDR ring, that are gravitationally bound and prone to collapse to form stars. We note that clumps that already have formed stars would appear to have α>1 as protostellar feedback tends to increase the virial mass. We emphasize that this calculation is only approximate and other realizations of a clump distribution are possible, such as a larger number of smaller and denser clumps.

Finally, we compare the observed line intensities with PDR model predictions using the KOSMA-τ PDR model *(81,82)*. The model solves the problem of physical (thermodynamic, excitation and radiative transfer) and chemical balance in a spherical model cloud and computes the resulting emissivities. These emissivities strongly depend on the physical and chemical structure of the PDR and on the degree of clumpiness within the PDR.

Using the KOSMA-τ model we find that the observed intensities of the red-shifted, blue-shifted and bulk components can originate exclusively from PDRs. The intensities are reasonably well reproduced by a non-clumpy model with densities between $10^3$ and $10^4$ cm$^{-3}$. These densities are lower than those expected from a shock at 15 km/s (see main text), possibly due to magnetic cushioning limiting compression behind the shock or from instabilities caused by the interaction of the hot plasma in the bubble with the shell. However, the densities agree with the absence of significant CO and [OI] 63µm emission. At densities of $10^5$ cm$^{-3}$ and higher these transitions would be considerably above the detection limit even assuming a significant beam filling effect.

The modeling results are uncertain because of the many uncertainties related to the complex structure of the source and the limited data at hand. We emphasize the need for more detailed simulations on the effects of compression in a gravitationally dominated regime.

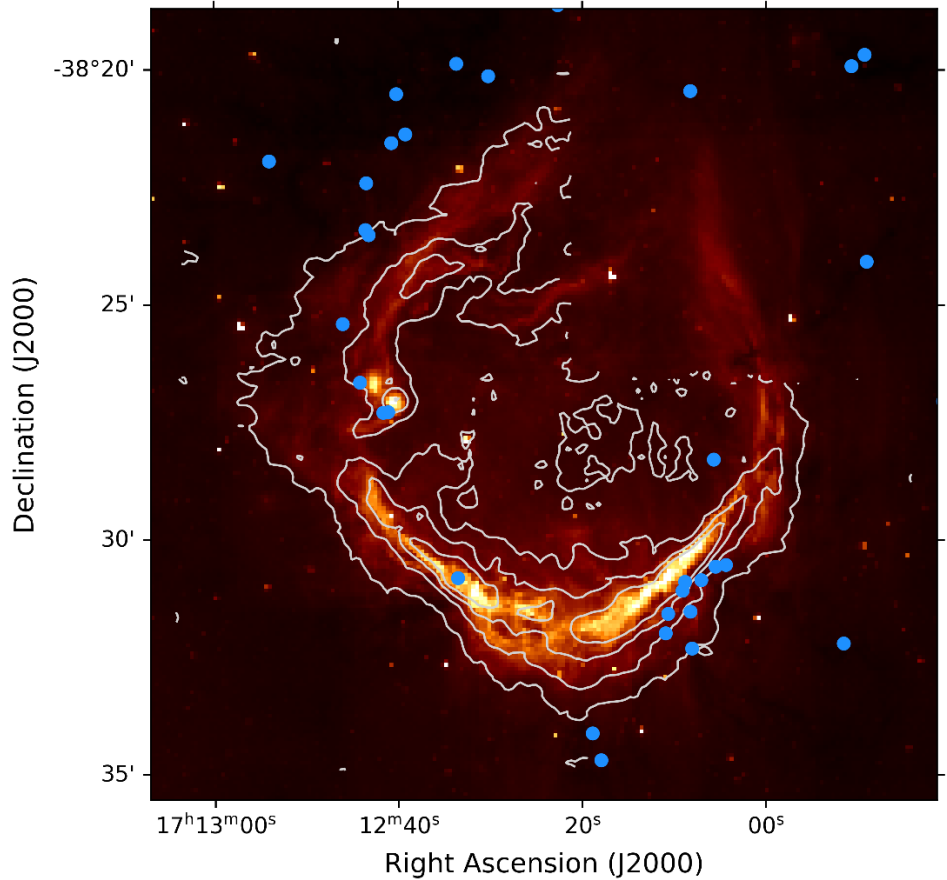

**Fig. S1.** Class I YSOs in the field of RCW 120 (blue dots), indicating the recent formation of stars. The YSOs are predominantly located in the dense ring surrounding RCW 120, suggesting that their formation was triggered by the expansion of the HII region *(15,16)*. The background image is of *Spitzer* GLIMPSE 8 μm emission and the white contours are of [CII] integrated intensity, scaled from 40 to 160 K km/s in 40 K km/s increments.

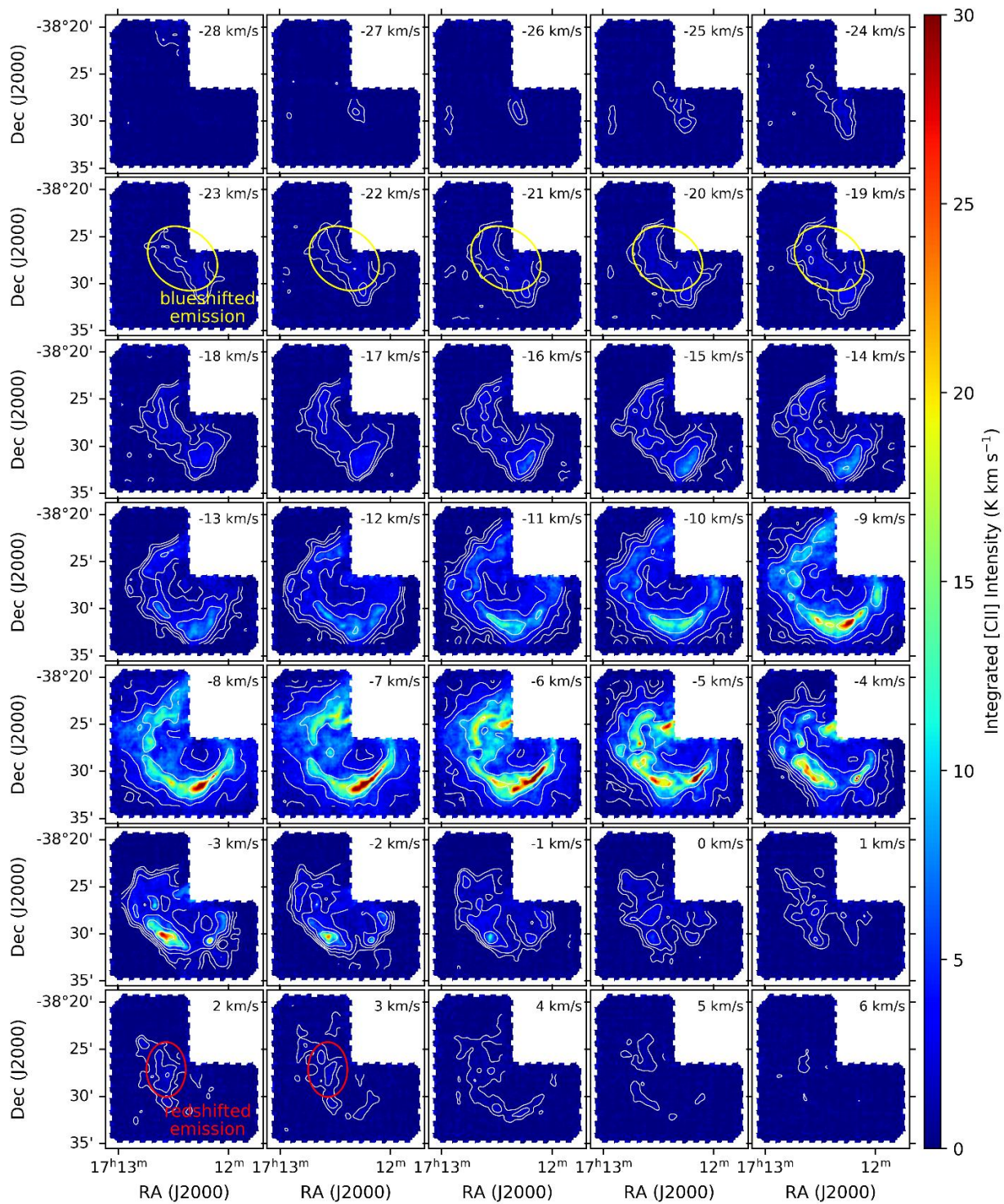

**Fig. S2.** [CII] integrated intensity maps. The maps are in 1 km/s intervals, centered at the velocity shown in the top right corner of each panel. Contours are at 0.5, 1, 2, 4, 8, and 16 K km/s. Blue-shifted emission from the interior of the region is clearly visible at lower velocities, indicating expansion toward us (marked with the yellow ellipses). Some red-shifted emission is also visible at positive velocities (marked with the red ellipses).

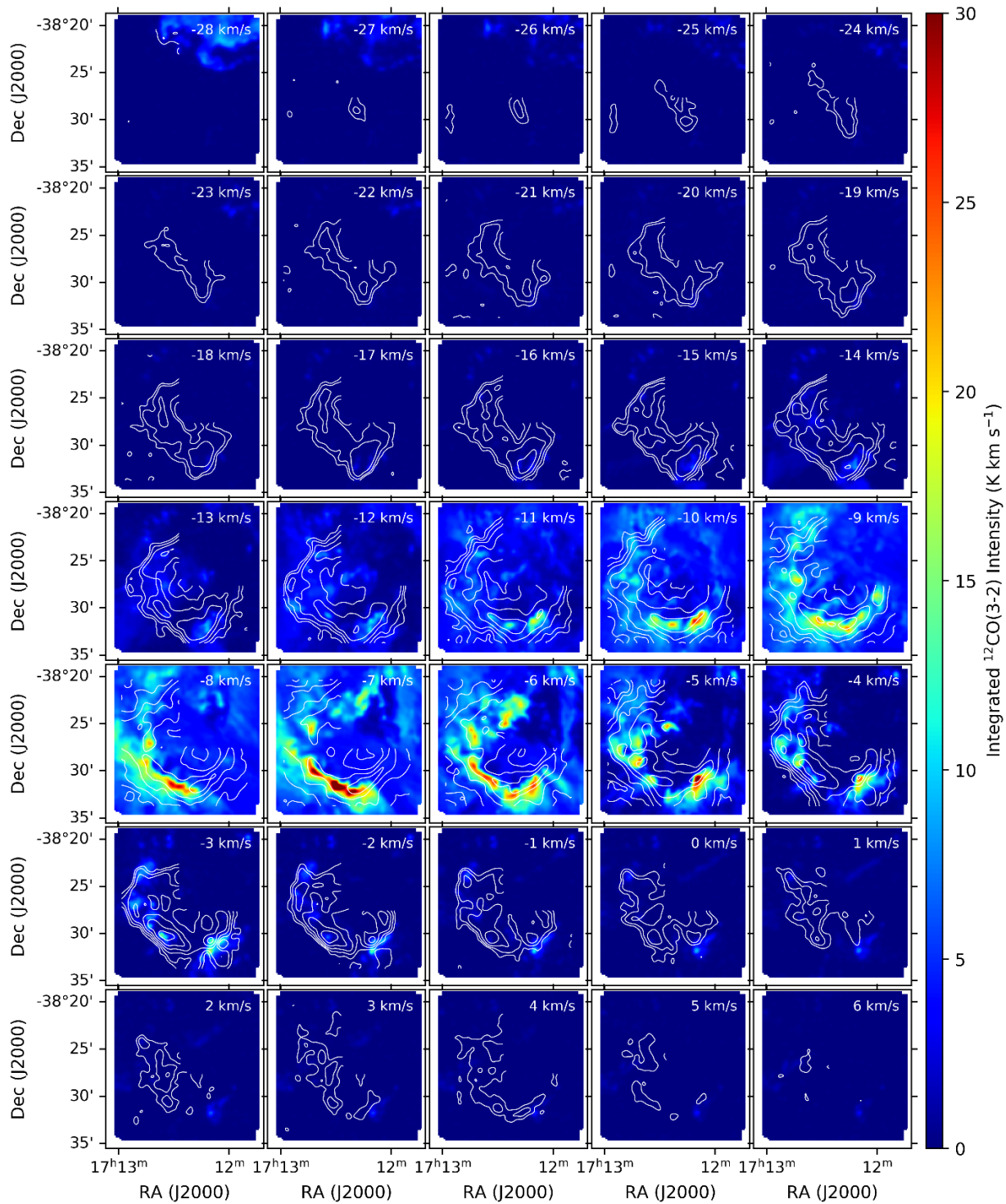

**Fig. S3.** $^{12}$CO(3-2) integrated intensity maps. The maps are in 1 km/s intervals, centered at the velocity shown in the top right corner of each panel. Contours are of [CII] at 0.5, 1, 2, 4, 8, and 16 K km/s. Most of the $^{12}$CO(3-2) emission is found at the systemic velocity of the region, although a separate molecular cloud is visible toward the north at low velocities. We do not detect an expansion signal in $^{12}$CO(3-2).

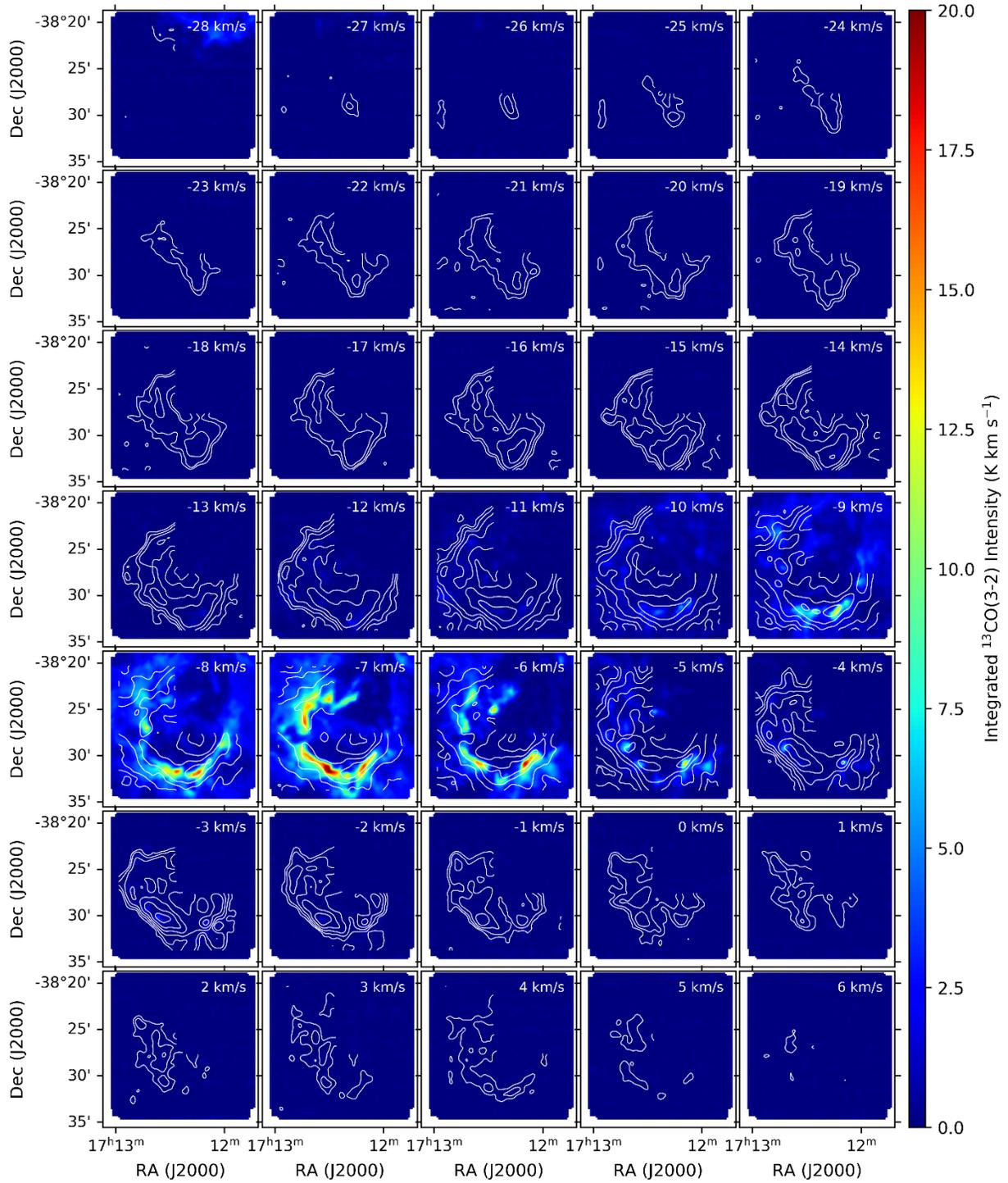

**Fig. S4.** $^{13}$CO(3-2) integrated intensity maps. The maps are in 1 km/s intervals, centered at the velocity shown in the top right corner of each panel. Contours are of [CII] at 0.5, 1, 2, 4, 8, and 16 K km/s. Essentially all the $^{13}$CO(3-2) emission is found near the systemic velocity of the region, with the exception of the separate molecular cloud to the north. Similar to the $^{12}$CO(3-2) data, we do not detect an expansion signal in $^{13}$CO(3-2).

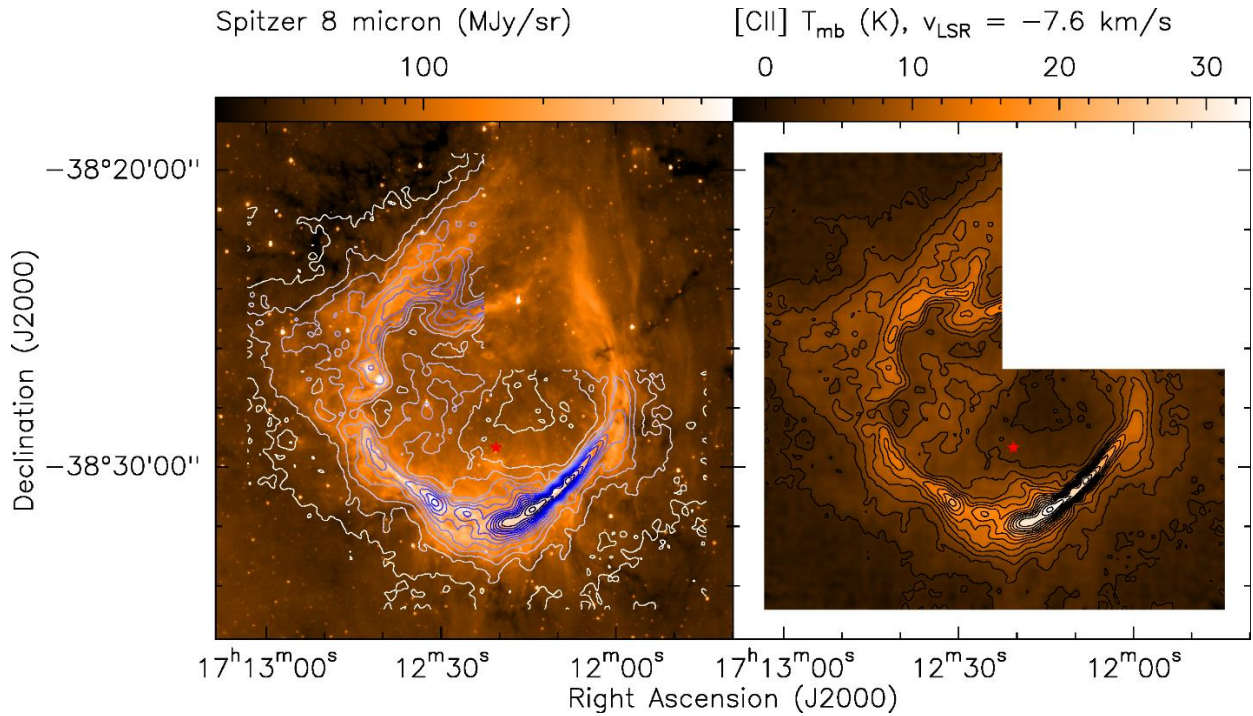

**Fig. S5.** Overview of the distribution and kinematics of 8 μm emission (left) and [CII] emission (right). The contours in the left panel are of [CII] emission. The distribution of the [CII] emission roughly follows that of 8 μm.

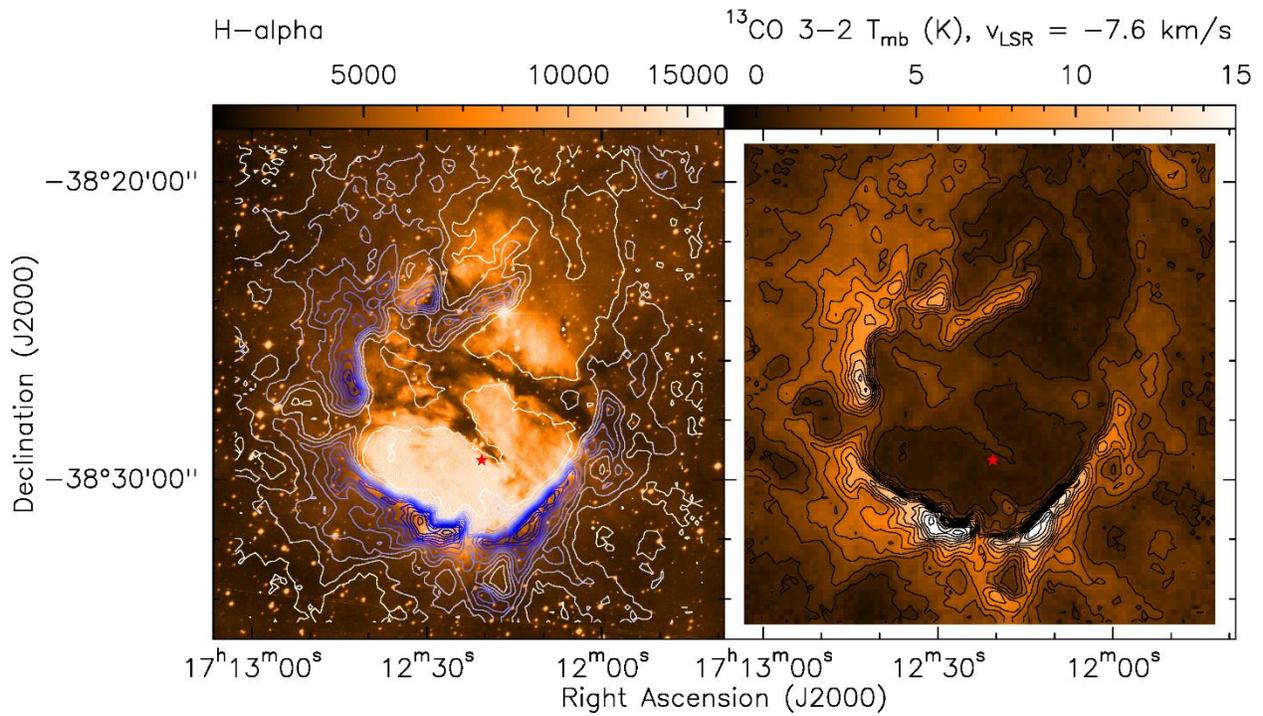

**Fig. S6.** Overview of the distribution and kinematics of Hα emission (left) *(83)* and APEX $^{13}$CO(3-2) emission (right). The contours in the left panel are of APEX $^{13}$CO emission. The foreground structure seen in Hα extinction is also visible in $^{13}$CO.

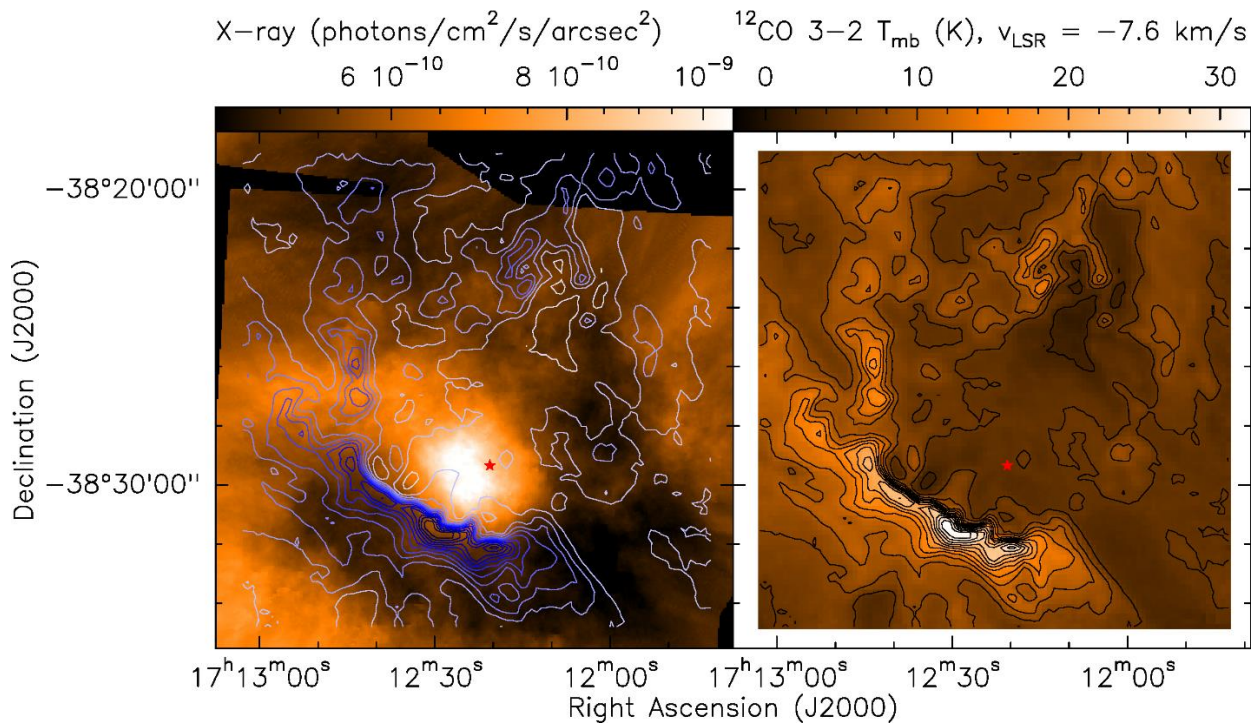

**Fig. S7.** Overview of the distribution and kinematics of diffuse X-ray emission (left) and APEX $^{12}$CO(3-2) emission (right). The contours in the left panel are of APEX $^{12}$CO emission. The hot plasma traced by the diffuse X-ray emission is breaking out of the shell toward the east and likely also toward the north, where the X-ray emission is attenuated by a foreground molecular cloud. The lack of $^{12}$CO emission toward the south-west is due to self-absorption and not due to the absence of molecular gas.

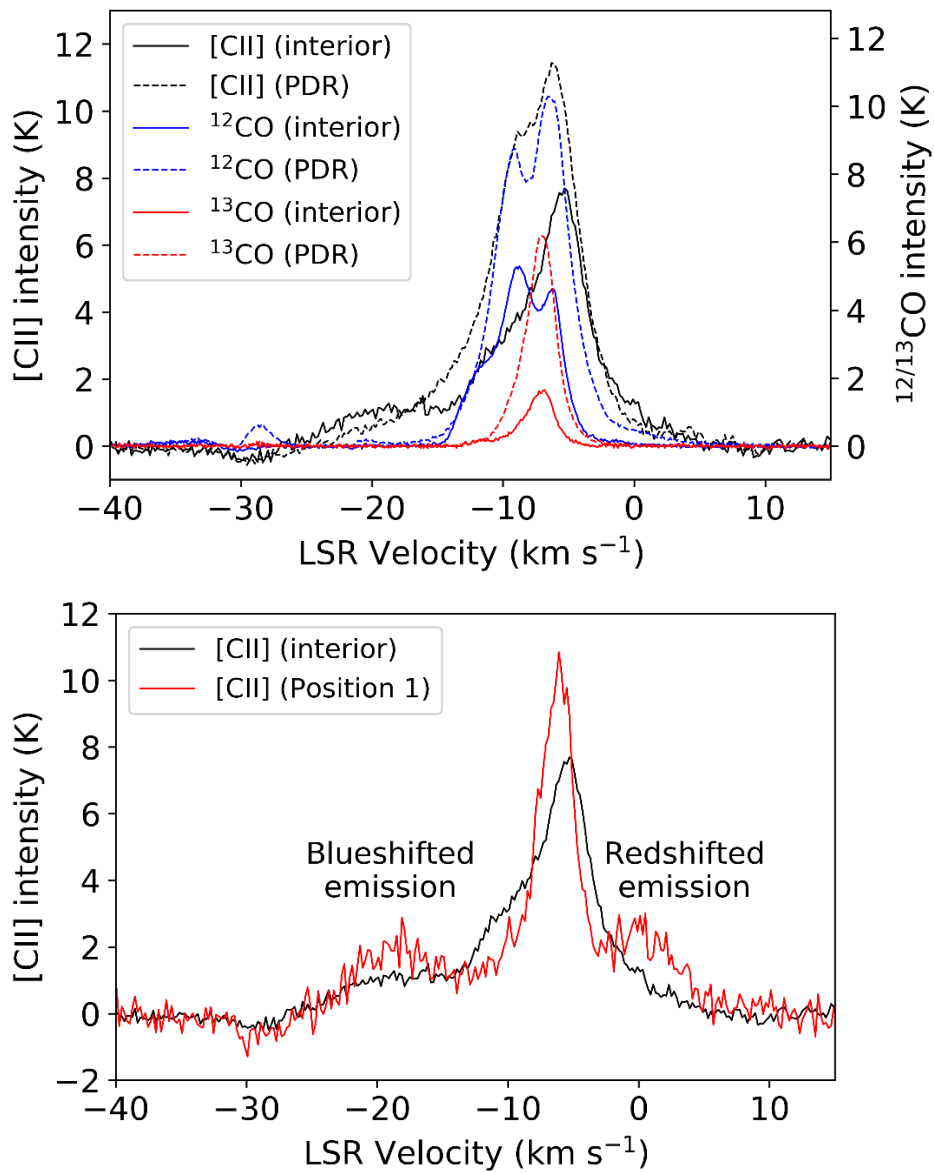

**Fig. S8.** Spectra of [CII], $^{12}$CO(3-2), and $^{13}$CO(3-2) toward RCW 120. Top: The spectra are averaged over the interior of the region (solid lines) and over the annular PDR shown in Figure 1 (dashed lines). Bottom: [CII] emission averaged over the interior of the region (black line) and toward RA, Dec [J2000] = $17^h12^m27.65^s$, -38°27'28.4" (labeled as "Position 1" and indicated in Figure 1). There is a blue-shifted component near ~ -20 km/s, and a red-shifted component near 0 km/s.

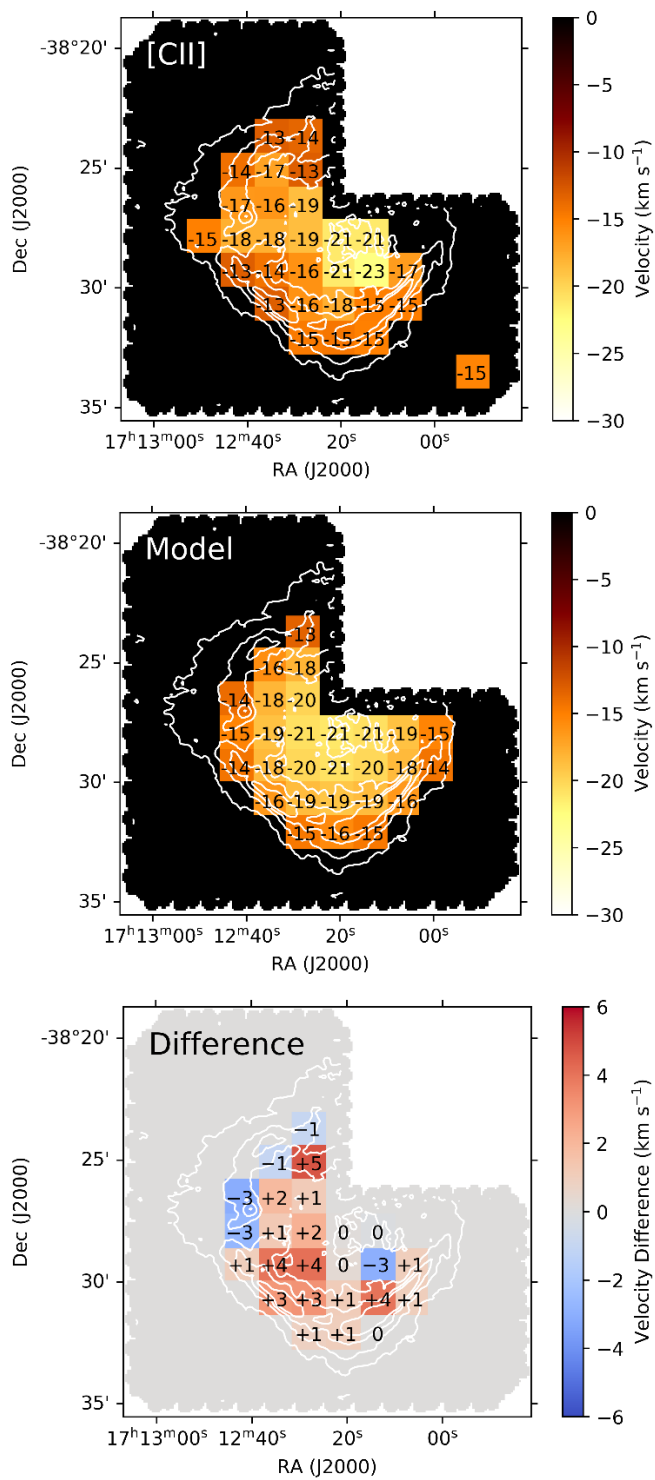

**Fig. S9.** Velocities of the blue-shifted shell expanding toward us. Top: Measured blue-shifted [CII] velocities. The contours are of integrated intensity. Middle: Velocities expected from an isotropically expanding half-shell with an expansion speed of 14 km/s, and a systemic velocity of -7.5 km/s. Bottom: Difference between the observed velocities and those from an isotropically expanding half-shell.

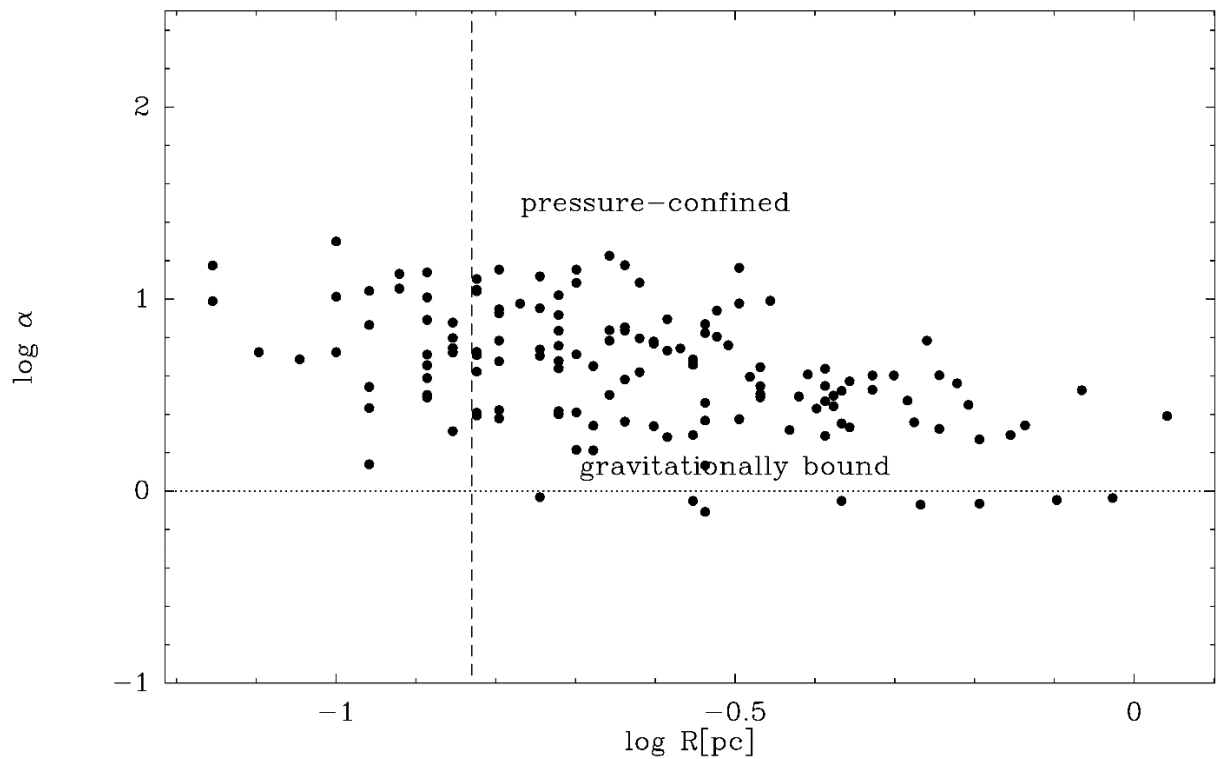

**Fig. S10.** Virial parameter α as a function of clump radius. While most clumps have log α > 0 and are pressure-confined or transient, several clumps—located mostly in the PDR—have log α < 0, indicating that they are gravitationally bound and prone to collapse to form stars. The dotted horizontal line indicates the transition between pressure-confined and gravitationally bound clumps, and the dashed vertical line corresponds to the beam width of the $^{13}$CO(3-2) data at the distance of RCW 120.

**Captions for Movies**

**Movie S1.** Animated version of Fig. S5.

**Movie S2.** Animated version of Fig. S6.

**Movie S3.** Animated version of Fig. S7.